\documentclass[%
preprint%
 ,secnumarabic%
,superscriptaddress%
,amssymb, amsmath,nofootinbib,aps]{revtex4-1}

\newcommand{\be}{\begin{equation}}
\newcommand{\ee}{\end{equation}}

 \newcommand{\bea}{\begin{eqnarray}}
\newcommand{\eea}{\end{eqnarray}}

\usepackage{epsfig}%
\usepackage{graphicx}%
\usepackage{color}
\usepackage[colorlinks=true,linkcolor=blue]{hyperref}%
\expandafter\ifx\csname package@font\endcsname\relax\else
 \expandafter\expandafter
 \expandafter\usepackage
 \expandafter\expandafter
 \expandafter{\csname package@font\endcsname}%
\fi

\begin{document}

\title{Stability constraints in triplet extension of MSSM}
\author{Moumita Das}
\email{moumita@prl.res.in}
\affiliation{Indian Statistical Institute, 203 Barrackpore Trunk Road, Kolkata 700108, India}
\author{Stefano Di Chiara}
\email{stefano.dichiara@helsinki.fi}
\affiliation{Helsinki Institute of Physics, P.O.Box 64, FI-00014, University of Helsinki, Finland}
\author{Sourov Roy}
\email{tpsr@iacs.res.in}
\affiliation{Department of Theoretical Physics, Indian Association for the Cultivation of Science, 
2A $\&$ 2B Raja S.C. Mullick Road, Kolkata 700032, India}

\def\be{\begin{equation}}
\def\ee{\end{equation}}
\def\al{\alpha}
\def\bea{\begin{eqnarray}}
\def\eea{\end{eqnarray}}


\begin{abstract}
We study the stability constraints on the parameter space of a triplet extension of MSSM. Existence of unbounded from below directions in the potential can spoil successful Electroweak (EW) symmetry breaking by making the corresponding minimum unstable, and hence the model should be free from those directions. Avoiding those directions restricts the parameter space of the model. We derive four stability constraints, of which only three independent from each other. After scanning the model's parameter space for phenomenologically viable data points, we impose the stability constraints and find that only about a quarter of the data points features a stable EW minimum. At those data points featuring stability, $\mu$ and the up Higgs soft mass turn out to be smaller than about a TeV in absolute value, which make the mass of the lightest chargino and neutralino smaller than about 700 GeV. Two relevant phenomenological consequences of lifting the unbounded from below directions are that the lightest Higgs boson decay rate to diphoton predicted by the triplet extension of MSSM generally features larger deviations from MSSM and fine tuning is actually higher, that what each of the two would be without imposing stability constraints.
\end{abstract}
\maketitle

\section{Introduction}
The Triplet Extended Supersymmetric Standard Model (TESSM) was introduced mainly to enhance the tree level Higgs boson mass while satisfying the top-quark mass bound 
\cite{Espinosa:1991wt,Espinosa:1991gr}. The authors have shown there that Electroweak (EW) symmetry breaking is 
successfully realized in this model, and studied constraints on the parameter space which must be satisfied 
for the potential to be stable. The lightest Higgs boson mass has been calculated in 
\cite{Espinosa:1991wt,Espinosa:1991gr,FelixBeltran:2002tb} for the Minimal Supersymmetric Standard 
Model (MSSM) with extra scalar triplet chiral superfields with hypercharge, respectively, ${\rm Y}=0,\pm 1$. The one 
loop correction to the Higgs boson mass was calculated for MSSM with a  ${\rm Y}=0$ scalar triplet also in 
\cite{DiChiara:2008rg}. The authors there pointed out that a light Higgs boson mass of $O(100)$~GeV can be generated already at tree level, if the triplet coupling to a pair of Higgs bosons is large and comparable to 
the top-quark Yukawa coupling. More recently it has been shown that the triplet charged states in 
TESSM can comfortably enhance the diphoton decay rate of the Higgs boson to match the value observed at LHC 
\cite{Delgado:2012sm,Kang:2013wm}. Further phenomenological studies of TESSM explored neutrino mass 
generation and leptogenesis \cite{Hambye:2000ui}, charged Higgs production at colliders \cite{DiazCruz:2007tf}, 
spontaneous CP violation \cite{Ham:2008ks}, etc. 

In this paper, we are interested in studying the stability of the EW minimum of the TESSM scalar potential. If the EW minimum is not a global minimum, correct EW symmetry breaking is not realized and its viability spoiled. \footnote{We assume throughout this paper the age of the Universe to be infinite.} It is therefore 
important to determine the constraints on the parameter space that ensure that the EW minimum is stable. For MSSM, there are a few directions possible along which the potential becomes 
Unbounded From Below (UFB), as it has been shown in \cite{Casas:1995pd}. The authors 
there also discussed the Charge and Colour Breaking (CCB) minima for MSSM. Other relevant studies on unstable 
and metastable minima of the supersymmetric scalar potential can be found in 
\cite{Datta:2000xy,Datta:2001dc,Gabrielli:2001py,Abel:1998ie,Blum:2009na,Chattopadhyay:2014gfa}. The problems associated with this kind of minima 
were addressed in \cite{Bajc:2011iu}. The study of unrealistic vacua or CCB minima in different 
supersymmetric models have already been discussed in \cite{Kanehata:2010ci} for MSSM with neutrino 
mass operators, in \cite{Kanehata:2011ei,Kobayashi:2012xv} for NMSSM, in \cite{Kobayashi:2010zx} 
for $\nu$SSM etc.

For TESSM, conditions for the stability of the potential have been derived in \cite{Espinosa:1991wt}. Indeed, as we have pointed out above, there may be a few unaccounted for UFB directions in the field space. Hence we would like to perform a full analysis of the UFB directions of the TESSM potential. We shall show that stability constraints that lift these directions allow one to constrain severely the parameter space of TESSM, with observable consequences for the mass spectrum and TESSM phenomenology.

This paper is structured as follows: In Section~\ref{model} we introduce the model, define the EW minimum of the potential, and discuss the present theoretical constraints. 
Next, we present the main results of this paper, which are the UFB directions of the TESSM potential and the corresponding stability constraints on the EW mininum, in Section~\ref{UFBD}. 
In the subsequent Section~\ref{ViaPS} we impose the stability constraints on a large set of data points, satisfying the present experimental constraints, and show how various 
relevant quantities are affected by the new constraints. Finally we draw the conclusions in Section~\ref{con}.


\section{The Model}\label{model}
The field content of TESSM is equal to that of MSSM extended by a $Y=0$ SU$(2)$ triplet chiral superfield, whose scalar component can be written in matrix form as
\be
T=\left(\begin{array}{cc}\frac{1}{\sqrt{2}} T^0 & T^+ \\T^- & -\frac{1}{\sqrt{2}}T^0\end{array}\right)\ .
\ee
The renormalizable superpontential of TESSM includes only two extra terms as compared to MSSM~\footnote{To simplify the phenomenology of the model we neglect the first and second generation 
Yukawas, as well as R-parity breaking terms.}, given that the cubic triplet term is zero:
\be\label{SP}
W_{\rm TESSM}=\mu_T {\rm Tr}(\hat T \hat T) +\mu_D \hat H_d\!\cdot\! \hat H_u + \lambda \hat H_d\!\cdot\! \hat T \hat H_u + y_t \hat U \hat H_u\!\cdot\! \hat Q - y_b \hat D \hat H_d\!\cdot\! \hat Q- y_\tau \hat E \hat H_d\!\cdot\! \hat L\ ,
\ee
where "$\cdot$" represents a contraction with the Levi-Civita symbol $\epsilon_{ij}$, with $\epsilon_{12}=-1$, and a hatted letter denotes a superfield. For later convenience we write explicitly the potential D and (quartic) F terms derived from the superpotential above:
\bea\label{VD}
V_{\rm D}&=&\frac{g_3^2}{2}\left[ \tilde{Q}_L^\dagger {\cal T}^a_3 \tilde{Q}_L-\tilde{u}_R {\cal T}^{*a}_3 \tilde{u}_R^*-\tilde{d}_R {\cal T}^{*a}_3 \tilde{d}_R^*\right]^2+\frac{g_Y^2}{2}\left[ \phi_j^\dagger Y \phi_j\right]^2\nonumber\\
&+&\frac{g_L^2}{2}\left[ H_u^\dagger {\cal T}^b_2 H_u+H_d^\dagger {\cal T}^b_2 H_d+{\rm Tr}\!\left(T^\dagger {\cal T}^b_2 T-T {\cal T}^b_2 T^\dagger \right)+\tilde{Q}_L^\dagger {\cal T}^b_2 \tilde{Q}_L+\tilde{L}_L^\dagger {\cal T}^b_2 \tilde{L}_L\right]^2\ ,
\eea
\be\label{VF}
V_{\rm F}=\left| \frac{\partial W_{\rm TESSM}}{\partial \phi^c_j} \right|^2\ ,
\ee
where $\cal T$ are group generators with gauge indices $a$ and $b$, and $\phi^c_j$ runs over the scalar components of all the TESSM chiral superfields.~\footnote{In Eqs.~(\ref{SP},\ref{VD}) as usual 
we dropped all the gauge and family indices that can be contracted, implied summation over repeated indices, and denoted with a tilde the scalar superpartner of a SM fermion field.} Also, $g_3$, $g_L$, and $g_Y$ are the $SU(3)_C$, $SU(2)_L$, and the $U(1)_Y$ gauge couplings, respectively.	
The full potential is then given by:
\be\label{Vtot}
V=V_{\rm D}+V_{\rm F}+V_S\ ,
\ee
where $V_S$ represents the soft supersymmetry breaking terms corresponding to the superpotential in Eq.~\eqref{SP}, as well as the soft squared masses associated to each scalar field:
\bea
V_S&=&\left[\mu_T B_T {\rm Tr}(T T) +\mu_D B_D H_d\!\cdot\! H_u + \lambda A_T H_d\!\cdot\! T H_u + y_t A_t \tilde{t}_R^* H_u\!\cdot\! \tilde{Q}_L + h.c.\right]  \nonumber\\
       & & + m_T^2 {\rm Tr}(T^\dagger T) + m_{H_u}^2 \left|H_u\right|^2  + m_{H_d}^2 \left|H_d\right|^2 + \ldots  \ .
\label{softV}\eea

To break correctly the EW symmetry, SU$(2)_L\times$U$(1)_Y\rightarrow$~U$(1)_{\rm EM}$, we assign nonzero vevs only to the neutral scalar components and impose the usual minimization conditions on 
the potential in Eq.~\eqref{Vtot}
\begin{gather}
H^0_u\equiv \frac{1}{\sqrt{2}}\left(a_u+i b_u  \right)\ , \quad H^0_d\equiv \frac{1}{\sqrt{2}}\left(a_d+i b_d  \right)\ , \quad T^0\equiv \frac{1}{\sqrt{2}}\left(a_T+i b_T  \right)\ ; \nonumber\\
\partial_{a_i} V|_{\rm vev}=0\  , \quad \langle a_i\rangle = v_i \ , \quad  i=u,d,T\ ,
\label{EWvs}\end{gather}
where $a_i$ and $b_i$ are both defined to be real. The conditions above allow one to determine three of the Lagrangian free parameters in terms of the other ones:
\bea\label{stabV}
m_{H_u}^2&=&-\mu _D^2-\frac{g_Y^2+g_L^2}{8}  \left(v_u^2-v_d^2\right)+B_D \mu _D \frac{v_d}{v_u}-\frac{\lambda^2}{4}  \left(v_d^2+v_T^2\right)+ \lambda v_T \left[\mu
   _D-\left(\frac{A_T}{2}+\mu _T\right)\frac{v_d}{v_u}\right]\ ,\nonumber\\
m_{H_d}^2&=&-\mu _D^2+\frac{g_Y^2+g_L^2}{8}  \left(v_u^2-v_d^2\right)+B_D \mu _D \frac{v_u}{v_d}-\frac{\lambda^2}{4} \left(v_u^2+v_T^2\right)+ \lambda v_T \left[\mu
   _D-\left(\frac{A_T}{2}+\mu _T\right)\frac{v_u}{v_d}\right]\ ,\nonumber\\
 m_T^2&=&-\frac{\lambda ^2}{4}  \left(v_d^2+v_u^2\right)-2 \mu _T \left(B_T+2 \mu _T\right)+ \lambda \left[\mu _D \frac{v_d^2+v_u^2}{2 v_T}-
   \left(\frac{A_T}{2}+\mu _T\right) \frac{v_d v_u}{v_T}\right]  \ .
\eea
By plugging Eqs.~\eqref{stabV} in Eq.~\eqref{Vtot} one derives the expression for the potential at the EW minimum:
\be\label{VEW}
V_{\rm EW}=-\frac{g_Y^2+g_L^2}{32}  \left(v_d^2-v_u^2\right)^2-\frac{\lambda^2}{8} \left[v_d^2 v_u^2+v_T^2 \left(v_d^2+v_u^2\right)\right]-\frac{\lambda  v_T}{4} \left[v_d  v_u\left(A_T +2\mu_T\right)-\left(v_d^2+v_u^2\right) \mu _D\right] .
\ee
The EW potential given above receives relevant corrections at one loop, which can in principle be minimized by choosing a suitable renormalization scale close to the 
heaviest masses in the particle spectrum. Contrarily to MSSM, though, it is not possible to solve the EW vevs in terms of the couplings and dimensional parameters of TESSM, 
and therefore in principle one should run the vevs by using the Callan Symanzik equation for the effective potential of a softly broken supersymmetric theory \cite{Martin:2001vx}. 
Analogously to the NMSSM case \cite{Kanehata:2010ci,Kanehata:2011ei}, though, we choose to simplify our analysis by evaluating $V_{\rm EW}$ at the EW scale $v_w=246$~GeV. 
In Section~\ref{ViaPS} we find that the stability constraints obtained by comparing unrealistic vacua with $V_{\rm EW}(v_w)$ are generally conservative. 

The first stability condition for successful EW symmetry breaking is obtained by requiring the trivial vacuum at the origin to be unstable. By taking all the vevs to be zero, the requirement that one of the eigenvalues of the neutral scalar squared mass matrix be negative is equivalent to imposing the condition
\be\label{trivSt}
B_D^2>\mu _D^2 \left(\frac{m_{H_d}^2}{\mu _D^2}+1\right) \left(\frac{m_{H_u}^2}{\mu _D^2}+1\right) .
\ee
When the condition above is satisfied, one can derive an important bound on the mass of the lightest neutral Higgs \cite{Espinosa:1991wt,Espinosa:1991gr}:
\be\label{mhbnd}
m^2_{h^0_1}\leq m_Z^2 \left( \cos^2{2\beta} + \frac{\lambda^2}{g_Y^2+g_L^2} \sin^2{2\beta} \right)\ ,\quad \tan\beta=\frac{v_u}{v_d}\ .
\ee
The result in Eq.~\eqref{mhbnd} shows the main advantage and motivation of TESSM over MSSM: for $\tan\beta$ close to one and a large $\lambda$ coupling it is in principle possible in TESSM to generate the experimentally measured light Higgs mass already at tree-level \cite{DiChiara:2008rg}, which would imply no or negligible Fine Tuning (FT) of the model.

Even when the constraint in Eq.~\eqref{mhbnd} is satisfied, for any given values of the free parameters there can be colour and electromagnetic charge breaking minima that are deeper than the EW one or even unbounded from below (UFB) directions in the potential: given that the latter possibility gives the tightest constraints on the MSSM parameter space \cite{Casas:1995pd}, by analogy in this work we focus our attention on UFB directions in the TESSM potential and the corresponding stability constraints, which we derive in the next section.


\section{Unbounded From Below Directions}\label{UFBD}
In softly broken supersymmetric models the potential is generally stable, given that the quartic terms are generated by the superpotential as well as by the gauge interactions (D-terms) and the 
supersymmetric tree-level potential is semidefinite positive. If the quartic terms cancel, though, the soft mass squared terms can eventually drive the potential to negative infinite values. Our aim in 
this section is to perform first a complete analysis of the possible UFB directions in the TESSM tree-level potential, Eq.~\eqref{Vtot}, and then to derive the corresponding stability constraints on the 
parameter space.

In general to find the deepest UFB direction of a supersymmetric theory in an $N$ field subspace, one solves the minimization conditions with respect to $N - 1$ fields, and then substitutes the solutions 
in the potential which turns out not to have quartic terms. The combinations of vevs that can do the trick are those that can cancel separately each D and F quartic terms, Eqs~(\ref{VD},\ref{VF}), given 
that each of these terms is semidefinite positive. A straightforward way to cancel the SU$(3)$ D-terms is to take the vevs of the left-handed (LH) and right-handed (RH) squarks to be the same. 
Analogously, we take the positive and negative charged triplet component vevs equal to each other, which cancels their SU$(2)$ D-terms, and we do the same for the charged doublet components of $H_u$ and $H_d$, which cancels their U$(1)_Y$ D-terms. To simplify our analysis we define also the LH and RH charged slepton vevs to be equal to each other.  Finally, the cubic F terms, being supersymmetric, should be zero when the quartic terms involving the same fields are zero. We avoid cubic F terms from the outset by choosing sneutrino and slepton nonzero vevs to belong to different generations. In such a case, the surviving terms in the potential involving a stau vev can be readily obtained by simply relabeling those involving a sbottom vev, \footnote{The stop vev turns out to be zero along UFB directions because of its Yukawa coupling, and therefore indeed only the sbottom labels need to be changed.} and for this reason in the present analysis we simply neglect the stau vev and derive results involving it directly from those involving the sbottom vev. The remaining cubic terms, breaking supersymmetry softly, give constraints that are less tight than those obtained from UFB directions \cite{Casas:1995pd}: 
for this reason in this paper we focus on the latter constraints.

In summary, the set of nonzero charged real vevs we work with is defined by:
\be\label{vevc}
\langle \tilde{t}_L\rangle=\langle \tilde{t}_R\rangle=\frac{v_{\tilde{t}}}{\sqrt{2}}\ ,\ \langle \tilde{b}_L\rangle=\alpha\langle \tilde{b}_R\rangle=\frac{v_{\tilde{b}}}{\sqrt{2}}\ ,\ \langle H^+_u\rangle=\langle H^-_d\rangle=\frac{v_{H^\pm}}{\sqrt{2}}\ ,\ \langle T^+\rangle=\langle T^-\rangle=\frac{v_{T^\pm}}{\sqrt{2}}\ ,
\ee
where we introduced a phase $\alpha=\pm1$ for later convenience, while the neutral ones are:
\be\label{vevn}
\langle \tilde{\nu}_L\rangle=\frac{v_{\tilde{\nu}}}{\sqrt{2}}\ ,\ \langle H^0_u\rangle=\frac{v_{H^0_u}}{\sqrt{2}}\ ,\ \langle H^0_d\rangle=\frac{v_{H^0_d}}{\sqrt{2}}\ ,\ \langle T^0\rangle=\frac{v_{T^0}}{\sqrt{2}}\ ,
\ee
where we used a labeling for the neutral vevs different from that in Eqs.~\eqref{VEW} to distinguish the vevs associated with the EW minimum from the unrealistic ones. Moreover, to simplify our following results without loss of generality, in the rest of the paper we assume $v_{H^0_u}$ and all the Yukawa couplings to be positive.

In the next subsection we determine which sets of vevs, among those defined in Eqs.~(\ref{vevc},\ref{vevn}), allow for a UFB direction in the potential.

\subsection{Relevant Vevs}
To determine the sets of nonzero vevs which allow for UFB directions, we look for those vevs combinations that can cancel all the D and quartic F terms. Assuming all the masses and couplings in Eq.~\eqref{SP} to be nonzero, requiring the superpotential derivative with respect to the triplet components in Eq.~\eqref{VF} to cancel, we obtain:
\be\label{SC1}
\frac{\partial W_{\rm TESSM}}{\partial \phi^c_j}=0\ ,\quad \phi^c_j=T^0,T^+,T^-\ \Rightarrow v_{H^\pm}=0\land \left(v_{H^0_d}=0\lor v_{H^0_u}=0\right)\ .
\ee
Besides  $v_{H^\pm}$, Eq.~\eqref{SC1} requires either $v_{H_u^0}$ or  $v_{H_d^0}$ to be zero. Indeed it can be shown that there is no vevs combination canceling all the quartic terms for nonzero  $v_{H_d^0}$, and therefore we impose:
\be\label{SC2}
v_{H_d^0}=0.
\ee 
Having defined $v_{H_u^0}$ to be nonzero, we notice that $m^2_{H_u^0}$, being large and negative at the EW scale for data points that feature a viable EW minimum, can generate a deep UFB direction. 

After imposing Eq.~\eqref{SC2} and requiring the cancellation of the quartic F terms corresponding to the $H^0_u$ and $H^-_d$ fields, also the stop and charged triplet vevs turn out to be zero:
\be\label{SC3}
\frac{\partial W_{\rm TESSM}}{\partial \phi^c_j}=0\ ,\quad \phi^c_j=H^0_u,H^-_d\ \Rightarrow v_{\tilde{t}}=v_{T^\pm}=0\ .
\ee
Having set to zero the charged doublet and triplet Higgs vevs as well as the stop and the neutral down Higgs ones according to Eqs.~(\ref{SC1},\ref{SC2},\ref{SC3}), the only nonzero D and quartic F terms left are, respectively,
\be
V_D\supset \frac{g_Y^2+g_L^2}{32}\left( v_{\tilde{b}}^2- v_{\tilde{\nu}}^2 + v_{H_u^0}^2\right)^2\ ,\quad V_F\supset\frac{1}{4} \left(y_b^2 v_{\tilde{b}}^4+\sqrt{2} y_b \alpha\lambda  v_{\tilde{b}}^2 v_{H^0_u} v_{T^0}+\frac{\lambda ^2}{2} v_{H^0_u}^2 v_{T_0}^2\right)\ .
\ee
Assuming a nonzero neutral up Higgs vev, we find therefore that it is possible to cancel all the quartic terms for the following sets of nonzero vevs:
\be\label{no4tvu}
v_{H^0_u}\neq 0\ \land\ v_{\tilde{\nu}}\neq 0\ \land \left(  \left( v_{T^0}= 0\ \land\ v_{\tilde{b}}= 0 \right)\lor \left( v_{T^0}= 0\ \land\ v_{\tilde{b}}\propto\sqrt{v_{H_u^0}} \right)  \lor  \left( v_{T^0}\neq 0\ \land\ v_{\tilde{b}}\neq 0 \right) \right)\ ,
\ee
with the other vevs being all identically zero. The only other possible set of non-trivial vevs canceling all the quartic terms is
\be\label{no4tvT}
v_{T^0}\neq 0\ \lor\ v_{T^\pm}\neq 0\ ,
\ee
with the other vevs, including $v_{H^0_u}$, being all identically zero. Evidently the potential in the UFB directions identified by Eq.~\eqref{no4tvT} has the same form for either of the two vevs and one can work just with $v_{T^0}$.

In the following subsection we work out the expressions for the potential along the four UFB directions expressed by Eqs.~(\ref{no4tvu},\ref{no4tvT}) and define the stability constraints associated with each of them.

\subsection{Stability Constraints}
We start with the simplest case, the UFB direction defined by Eq.~\eqref{no4tvT}: by setting to zero every vev but $v_{T^0}$ in the potential (the corresponding result for $v_{T^\pm}$ is the same), Eq.~\eqref{Vtot}, we get
\be\label{vUFB1}
V_{\rm UFB-1}=\frac{v_{T^0}^2}{2}  \left[m_T^2+2 \mu _T \left(B_T+2 \mu _T\right)\right]\ .
\ee
We evaluate Eq.~\eqref{vUFB1} at a renormalization scale $\Lambda$ of the order of the heaviest mass in the physical particle spectrum \cite{Casas:1995pd}, as to minimize the contribution of quantum corrections, which we neglect entirely. For the potential above the largest mass is roughly equal to $v_{T^0}$ multiplied by the largest of its couplings, generally either $g_L$ or $\lambda$. Rather than simply requiring the coefficient of the squared vev to be positive, a given point in parameter space is stable against tunneling \footnote{In this analysis we simplify the stability constraints by assuming the age of the universe to be infinite. The numerical factor in Eq.~\eqref{UFB1} agrees with \cite{Casas:1995pd}, though for the numerical analysis it has little relevance.} to UFB--1 if
\be\label{UFB1}
V_{\rm EW}\left(v_w\right)<V_{\rm UFB-1}\left(\Lambda\right)\ ,\quad v_w\leq\Lambda\leq\Lambda_{\rm UV}\ ,\quad v^2_{T^0}\sim2 \max\left[g^2_L,\lambda^2 \right]^{-1}\Lambda^2\ ,
\ee
where all the couplings and dimensionful parameters are evaluated at $\Lambda$. For this purpose we calculated the full set of beta functions, including those of the dimensionful parameters which, to the best of our knowledge, were not given in the literature and that we present in Appendix \ref{apbeta}. In Section~\ref{ViaPS} we elucidate how to implement in practice the stability constraint in Eq.~\eqref{UFB1} and those that follow in this Section.

Next we take up the slightly more complicated case with only two nonzero vevs, the first one in Eq.~\eqref{no4tvu}:
\be\label{V2}
V|_{v_{H^0_u}\neq 0,v_{\tilde{\nu}}\neq 0}=\frac{1}{2} \left[m_L^2 v_{\tilde{\nu }}^2+\left(m_{H_u^0}^2+\mu _D^2\right) v_{H_u^0}^2+\frac{g_Y^2+g_L^2}{16} \left(v_{\tilde{\nu }}^2-v_{H_u^0}^2\right)^2\right]\ .
\ee
where $m^2_L$ is the soft mass squared of the slepton doublet $L$. By requiring the potential above to be flat along the $v_{\tilde{\nu}}$ direction we find
\be
\partial_{v_{\tilde{\nu}}}V|_{v_{H^0_u}\neq 0,v_{\tilde{\nu}}\neq 0}=0\quad \Rightarrow v_{\tilde{\nu }}^2=v_{H_u^0}^2-\frac{8 m_L^2}{g_Y^2+g_L^2} \ ,
\ee
which, upon substitution in Eq.~\eqref{V2}, gives the deepest UFB direction corresponding to our choice of nonzero vacua
\be\label{vUFB2}
V_{\rm UFB-2}=\frac{v_{H_u^0}^2}{2}  \left(m_L^2+m_{H_u^0}^2+\mu _D^2\right)-\frac{2 m_L^4}{g_Y^2+g_L^2}\ .
\ee
Analogously to the stability constraint derived from the UFB--1 direction, a point in the TESSM parameter space may feature a stable EW minimum only if
\be\label{UFB2}
V_{\rm EW}\left(v_w\right)<V_{\rm UFB-2}\left(\Lambda\right)\ ,\quad v_w\leq\Lambda\leq\Lambda_{\rm UV}\ ,\quad v_{\tilde{\nu}}^2>0\ ,\quad v^2_{H_u^0}\sim2 \max\left[g^2_L,\lambda^2,y_t^2 \right]^{-1}\Lambda^2\ ,
\ee
where all the couplings and dimensionful parameters are evaluated at $\Lambda$.

The case with three nonzero vevs, the second one in Eq.~\eqref{no4tvu}, is a little more complicated, but the potential can be readily simplified by imposing its derivative with respect to $v_{\tilde{\nu}}$ to be zero:
\bea
\partial_{v_{\tilde{\nu}}}V|_{v_{H^0_u}\neq 0,v_{\tilde{b}}\neq 0,v_{\tilde{\nu}}\neq 0} &=& 0\quad \Rightarrow\quad v^2_{\tilde{\nu }}=v_{H_u^0}^2+v_{\tilde{b}}^2-\frac{8 m_L^2}{g_Y^2+g_L^2} \ ,\nonumber\\
V|_{v_{H^0_u}\neq 0,v_{\tilde{b}}\neq 0,v_{\tilde{\nu}}\neq 0} &=&\frac{y_b^2}{4} v_{\tilde{b}}^4+ \frac{v_{\tilde{b}}^2}{2} \left(m_L^2+m_Q^2+m_{\tilde{b}}^2-\sqrt{2} y_b \alpha  v_{H_u^0} \mu _D\right)+\frac{v_{H_u^0}^2}{2} \left(m_L^2+m_{H_u^0}^2+\mu _D^2\right)\nonumber\\
&-&\frac{2 m_L^4}{g_Y^2+g_L^2}\ .
\label{vUFB3a}\eea
By requiring the potential above to be flat along the $v_{\tilde{b}}$ direction and solving for $v_{\tilde{b}}$ we find
\be\label{vUFB3b}
\quad v_{\tilde{b}}^2=\frac{\sqrt{2} v_{H_u^0} y_b \left|\mu _D\right|-m_L^2-m_Q^2-m_{\tilde{b}}^2}{y^2_b}\ ,
\ee
where, with our assumption that $v_{H_u^0}$ and all the Yukawa couplings are positive, we simply took $\alpha$ equal to the sign of $\mu_D$. In turn the vevs in Eqs.~(\ref{vUFB3a},\ref{vUFB3b}) identify the deepest UFB direction corresponding to our choice of nonzero vacua
\bea\label{vUFB3}
V_{\rm UFB-3}&=& \left(m_L^2+m_{H_u^0}^2\right)\frac{v_{H_u^0}^2}{2}+\frac{\left|\mu _D\right|
   \left(m_L^2+m_Q^2+m_{\tilde{b}}^2\right) v_{H_u^0}}{\sqrt{2} y_b}-\frac{\left(m_L^2+m_Q^2+m_{\tilde{b}}^2\right)^2}{4 y_b^2}\nonumber\\
   &-&\frac{2 m_L^4}{g_Y^2+g_L^2}\ .
\eea
where $m^2_Q$ is the soft mass squared of the squark doublet $Q$. The corresponding stability constraint that any point in the TESSM parameter space has to satisfy is
\be\label{UFB3}
V_{\rm EW}\left(v_w\right)<V_{\rm UFB-3}\left(\Lambda\right)\ ,\quad v_w\leq\Lambda\leq\Lambda_{\rm UV}\ ,\quad v_{\tilde{\nu}}^2,v_{\tilde{b}}^2>0\ ,\quad v^2_{H_u^0}\sim2 \max\left[g^2_L,\lambda^2,y_t^2 \right]^{-1}\Lambda^2\ .
\ee
Notice that $V_{\rm UFB-3}$ is equal to the corresponding result in \cite{Casas:1995pd}, plus the second from last term in Eq.~\eqref{vUFB3}, which is large and negative: this is because the authors 
in \cite{Casas:1995pd} determined $v_{\tilde{b}}$ by requiring the $H_d^0$ quartic $F$ term to cancel, rather than by solving a minimization condition. In case $v_{H^0_u}$ is small, $v_{\tilde{b}}$ turns out to be imaginary, and so we can require the potential in Eq.~\eqref{vUFB3a} to be flat along the $v_{H^0_u}$ direction instead, which implies
\be\label{vUFB3c}
\quad v_{\tilde{b}}^2=\frac{\sqrt{2} v_{H_u^0}}{y_b} \left|\frac{m_L^2+m_{H_u^0}^2+\mu _D^2}{\mu _D}\right|\ .
\ee
In this case the potential along the deepest UFB direction is
\bea\label{vUFB3p}
V'_{\rm UFB-3}&=&\frac{\left(m_L^2+m_{H_u^0}^2\right) \left(m_L^2+m_{H_u^0}^2+\mu _D^2\right)}{2 \mu _D^2} v_{H_u^0}^2+\left|\frac{m_L^2+m_{H_u^0}^2+\mu _D^2}{\mu _D}\right| \frac{m_L^2+m_Q^2+m_{\tilde{b}}^2}{\sqrt{2}
   y_b} v_{H_u^0}\nonumber\\
   &-&\frac{2 m_L^4}{g_Y^2+g_L^2}\ ,
\eea
and the corresponding stability constraint reads the same as that in Eq.~\eqref{UFB3} but with $V'_{\rm UFB-3}$ replacing $V_{\rm UFB-3}$. Notice that, contrarily to the vev in Eq.~\eqref{vUFB3b}, the one in Eq.~\eqref{vUFB3c} is not generally smaller than $v_{H^0_u}$, and so one might be underestimating the heaviest mass, which is of the same order of the renormalization scale.  For viable points, though, $m_{H_u^0}^2$ is generally negative, in which case $v_{\tilde{b}}$ turns out to be of the order of $v_{H^0_u}$ or smaller. For this reason we still determine $v_{H^0_u}$ according to the last one in Eqs.~\eqref{UFB3} and then $v_{\tilde{b}}$ by Eq.~\eqref{vUFB3b}.

Finally, we take up the last scenario in Eq.~\eqref{no4tvu}: for $v_{H^0_u}\neq 0\land v_{\tilde{\nu}}\neq 0 \land  v_{T^0}\neq 0 \land v_{\tilde{b}}\neq 0,$ the requirement for the potential to be flat along the $\nu$ direction determines $v_{\tilde{\nu}}$ as given by Eq.~\eqref{vUFB3a}. The remaining minimization conditions produce rather involved solutions, which turn out to be complex on a large and disconnected region of field space. For this reason we choose simply to cancel the quartic F terms, which is achieved by setting
\be\label{vb3}
v_{\tilde{b}}^2=\frac{|\lambda v_{T^0} | v_{H^0_u}}{\sqrt{2}y_b}\ .
\ee

The potential along the plane identified by Eqs.~(\ref{vUFB3a},\ref{vb3}) expressed in terms of the remaining two vevs is
\bea\label{vUFB4}
V_{\rm UFB-4}&=&\frac{v_{T^0}^2}{2}  \left[m_T^2+2 \mu _T \left(B_T+2 \mu _T\right)\right]+\frac{v_{H_u^0}^2}{2}  \left(m_L^2+m_{H_u^0}^2+\mu _D^2\right)-\frac{2 m_L^4}{g_Y^2+g_L^2}\nonumber\\
&+&\frac{\left|\lambda  v_T\right| v_{H_u^0}}{2 \sqrt{2} y_b}\left(m_L^2+m_Q^2+m_{\tilde{b}}^2\right)\ .
\eea
By comparing the first line in the equation above with the first two UFB directions, Eqs.~(\ref{vUFB1},\ref{vUFB2}), and then realizing that the term in the second line is generally positive for 
phenomenologically viable data points, it is clear that any viable data point satisfying the first two stability constraints, Eqs.~(\ref{UFB1},\ref{UFB2}), is stable against tunneling to a vacuum along 
the plane defined by Eq.~\eqref{vUFB4}. \footnote{Notice that the last term in Eq.~\eqref{vUFB4} is not allowed to turn large and negative upon renormalization for points featuring a viable EW vacuum, 
because otherwise the stability constraint in Eq.~\eqref{UFB3} would not be satisfied.} For this reason in this analysis we disregard the UFB--4 stability constraint.

In the next section we test the stability of TESSM at data points that satisfy the most relevant phenomenological constraints from experiment. 


\section{Phenomenologically Viable Parameter Space}\label{ViaPS}

The first relevant phenomenological constraint is given by the $T$ parameter \cite{Peskin:1991sw}, which in TESSM receives a nonzero contribution already at tree level \cite{Espinosa:1991wt,Espinosa:1991gr}
\be
\alpha_e T=\frac{\delta m_W^2}{m_W^2}=\frac{4 v_T^2}{v^2}\leq 0.2\ ,\quad v^2=v_u^2+v_d^2\ ,\quad v_w^2=v^2+4 v_T^2=246^2~ {\rm GeV}^2
\ee
where $\alpha_e$ is the fine structure constant, the experimental constraint is at 95\%CL \cite{Beringer:1900zz}, and the vevs appearing in the expression above are those defined in Eqs.~\eqref{EWvs}. To satisfy the constraint above we take a small but non-zero fixed value for $v_T$
\be
v_T= 3\sqrt{2} ~{\rm GeV}\ .
\ee
We then scan a large region of the TESSM parameter space, defined by
\bea\label{pscan}
&&1\leq t_{\beta }\leq 10\ ,\ 5 \,\text{GeV}\leq \left|\mu _D,\mu _T\right|\leq 2 \,\text{TeV}\ ,\ 50 \,\text{GeV}\leq \left|M_1,M_2\right|\leq 1  \,\text{TeV}\ ,\nonumber\\ 
&& \left| A_t,A_T,B_D,B_T\right|\leq 2 \,\text{TeV}\ ,\ 500 \,\text{GeV}\leq m_Q,m_{\tilde{t}},m_{\tilde{b}}\leq 2 \,\text{TeV}\ ,
\eea 
for data points producing the observed mass spectrum for SM fermions and gauge bosons and satisfying the direct search constraints defined below
\bea
m_{h_1^0}=125.5\pm 0.1\, {\rm GeV}\ ;\ m_{A_{1,2}},\ m_{\chi^0_{1,2,3,4,5}}&\geq & 65\,{\rm GeV}\ ;\nonumber\\
m_{h^0_{2,3}} , m_{h^\pm_{1,2,3}}, m_{\chi^\pm_{1,2,3}}\geq 100\,{\rm GeV} \ ;\ m_{\tilde{t}_{1,2}},m_{\tilde{b}_{1,2}}&\geq & 650\,{\rm GeV}\ ,
\eea
where the neutral scalar masses are calculated by using the TESSM complete one loop effective potential \cite{DiChiara:2008rg}, while the others are tree level masses. The above bounds on the mass of pseudoscalars and neutralinos are actually tighter than the experimental ones \cite{Beringer:1900zz}, as to avoid the phenomenological complications of invisible decays of the light Higgs, which are though relevant for dark matter \cite{Arina:2014xya}. For each of the scanned data points the couplings retain perturbativity at least up to $10^4$ TeV, where we define a coupling constant to be perturbative if its renormalized value (obtained by running the given coupling by two loop beta functions \cite{Bandyopadhyay:2014tha}) is smaller than $2\pi$. \footnote{We choose 2$\pi$ because at larger values the couplings reach a fixed point which is an artifact of the series being truncated after the two loop contribution.} Moreover the experimental constraints on a heavy neutral Higgs mass is imposed by rescaling the CMS bound on the mass of a SM heavy Higgs decaying to $ZZ$, $m_{H^0}>770$ GeV \cite{CMS:2013ada}. 

After collecting 10957 viable data points satisfying the constraints outlined above, we test further their viability by imposing the stability constraints derived in the Section \ref{UFBD}.

To perform the stability test we pick 100 energies equally spaced on a logarithmic scale between the EW scale, 246 GeV, and the chosen UV scale, $\Lambda_{\rm UV}=10^4$ TeV, and run all the couplings and dimensional parameters to the 100 renormalization scales, with initial conditions defined by the given viable data point in the TESSM parameter space. For each of the 10957 viable data points we then evaluate the potential at the EW vacuum, Eq.~\eqref{VEW}, as well as at each of the 100 values along each of the UFB--1,2,3 directions, plus the one obtained from UFB--3, by replacing sbottom masses and couplings with the corresponding slepton quantities, which we call UFB--3$'$. The (real) vevs are automatically determined as functions of the given scale for each UFB direction. It turns out that only 24\% of the 10957 viable data points satisfy the stability constraints defined in Eqs.~(\ref{UFB1},\ref{UFB2},\ref{UFB3}). By applying each stability constraint individually one can assess the tightness of the constraints relative to each other: 58\% of points are stable against tunnelling to vacua along the UFB--1 direction, 95\% against UFB--2, 67\% against UFB--3, and just 41\% of the scanned data points feature an EW minimum deeper than vacua along the UFB--3$'$ direction, for scales up to $10^4$ TeV. Evidently the UFB--3$'$ constraint, defined by Eq.~\eqref{UFB3} with slepton masses and couplings replacing the corresponding sbottom quantities, is by far the tightest among the 3+1 constraints we imposed. This result is analogous to the one obtained for MSSM \cite{Casas:1995pd}. We also calculated the one loop potential evaluated at the EW minimum, and found it to be shallower than the tree level potential in Eq.~\eqref{VEW} for 76\% of data points, which makes the stability constraints defined by Eqs.~(\ref{UFB1},\ref{UFB2},\ref{UFB3}) generally conservative.

\begin{figure}[htb]
\centering
\includegraphics[width=0.46\textwidth]{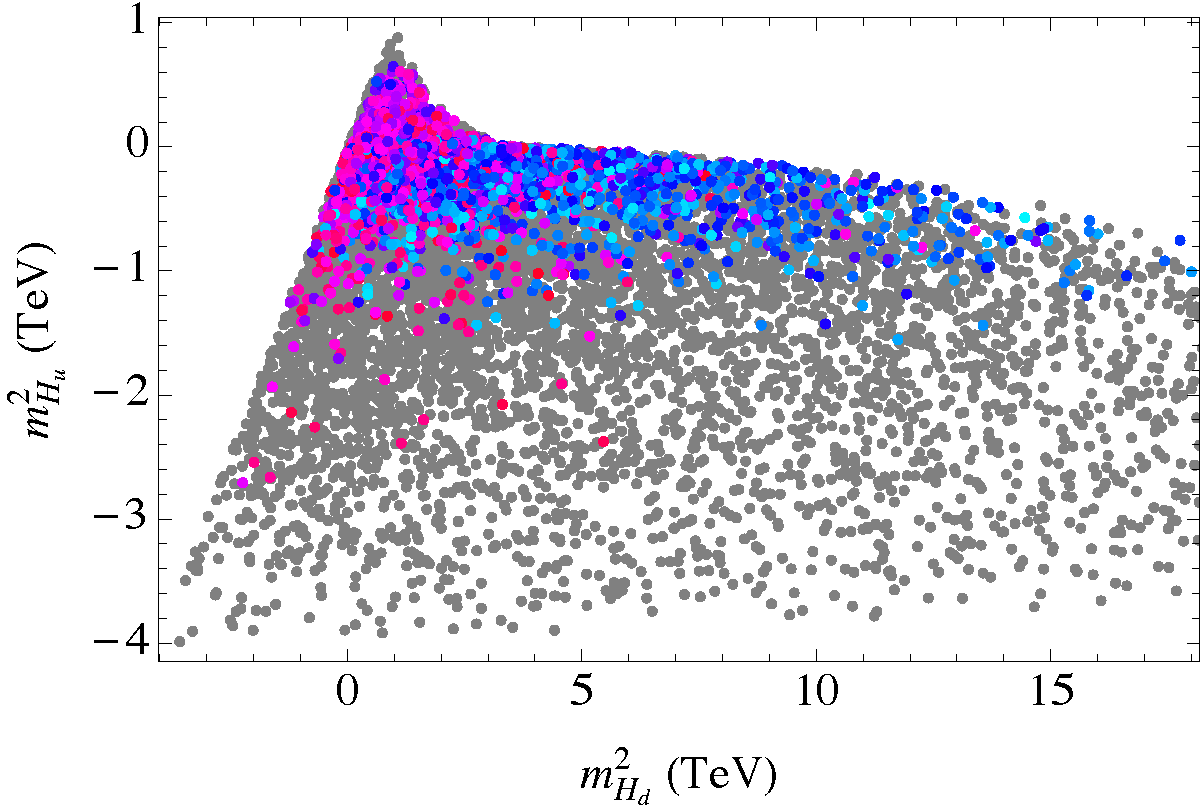}\hspace{0.1cm}
\includegraphics[width=0.12\textwidth]{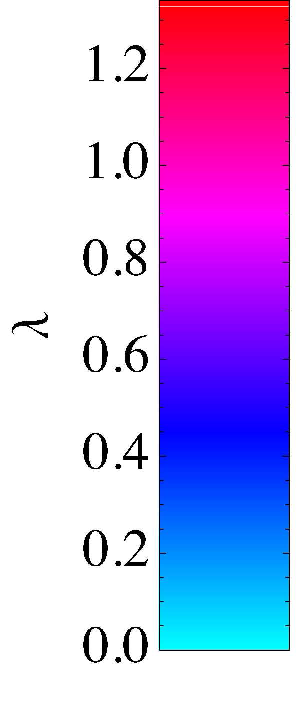}\hspace{0.1cm}
\caption{Scanned TESSM data points satisfying all phenomenological constraints elucidated in the Section~\ref{ViaPS}, projected on the $m^2_{H_d}-m^2_{H_u}$ plane (left panel). The grey points do not satisfy at least one of the stability constraints in Eqs.~(\ref{UFB1},\ref{UFB2},\ref{UFB3}), while the viable data points are colored according to the value of the triplet coupling $\lambda$ as shown in the right panel.}
\label{stabps}
\end{figure}

\begin{figure}[htb]
\centering
\includegraphics[width=0.46\textwidth]{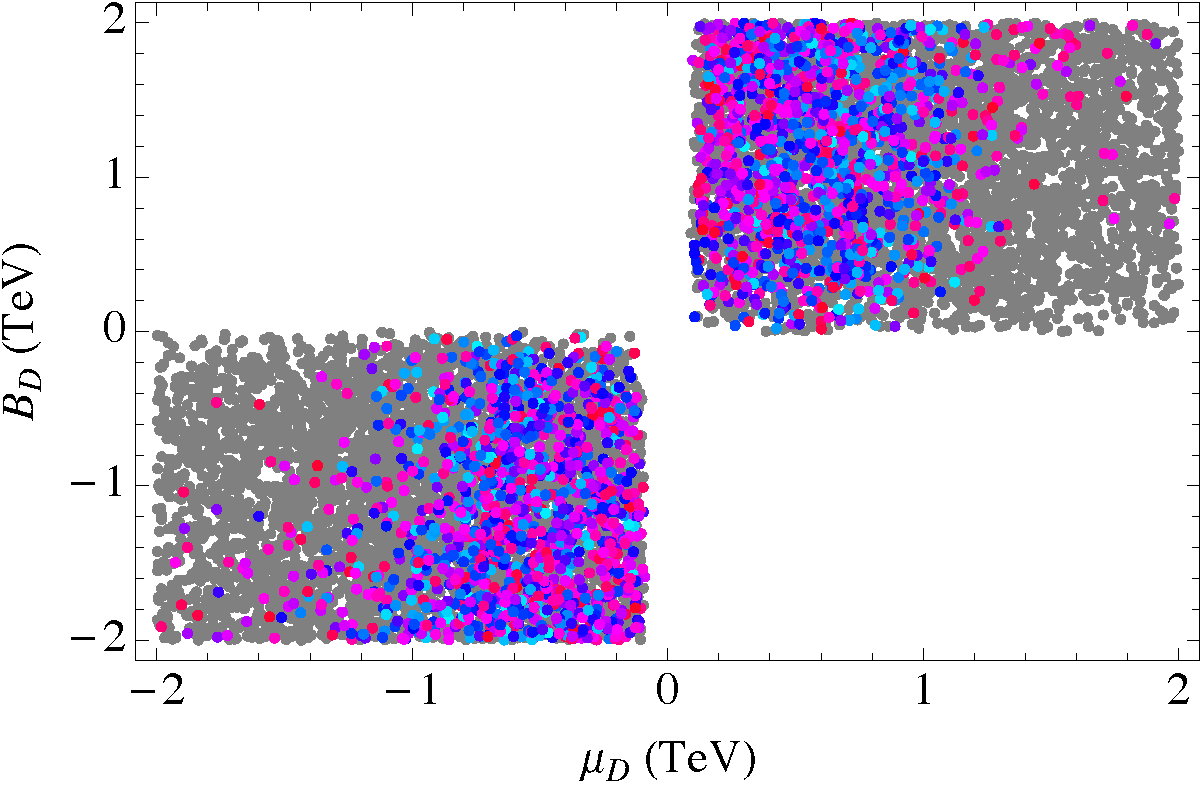}\hspace{0.1cm}
\includegraphics[width=0.46\textwidth]{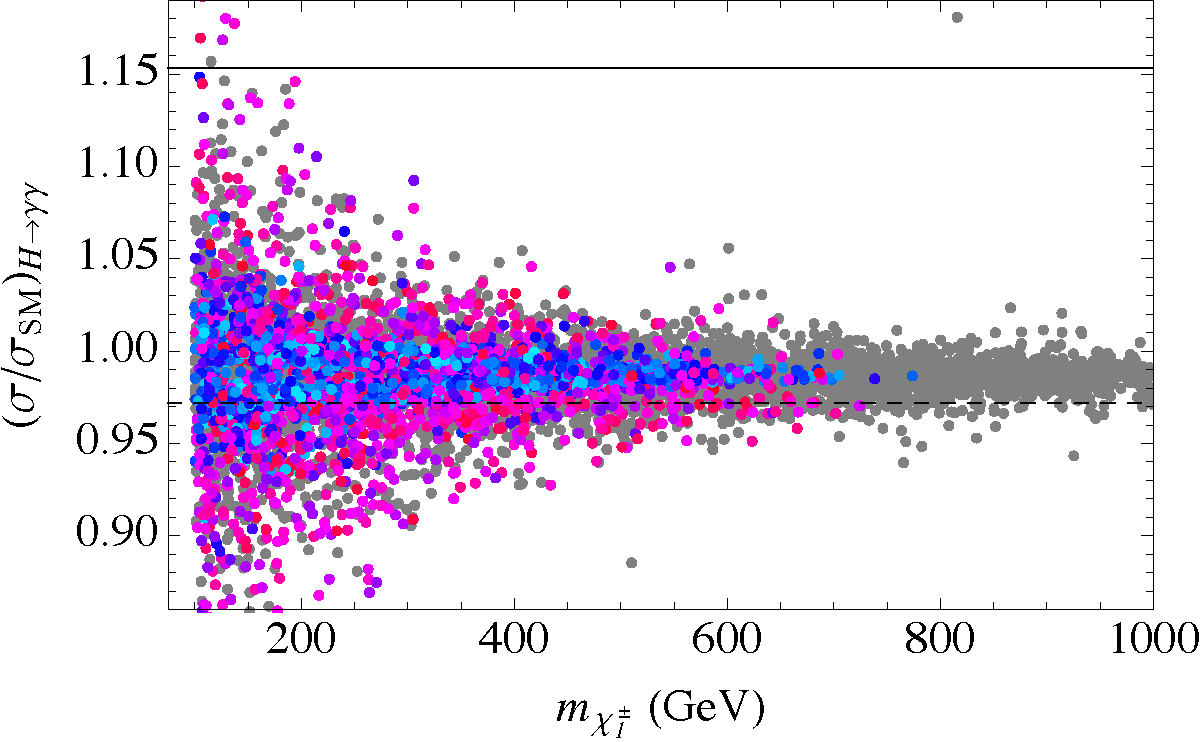}\hspace{0.1cm}
\caption{Scanned TESSM data points satisfying all phenomenological constraints elucidated in the Section~\ref{ViaPS}, projected on the $\mu_D-B_D$ plane (right panel), and Higgs decay rate to diphoton as 
a function of the mass of the lightest chargino (right panel). The grey points do not satisfy at least one of the stability constraints in Eqs.~(\ref{UFB1},\ref{UFB2},\ref{UFB3}), while the viable data points are colored according to the value of the triplet coupling $\lambda$ as shown in Fig.~\eqref{stabps} (right panel).}
\label{mchHgg}
\end{figure}

In Figs.~(\ref{stabps},\ref{mchHgg}) we plot all the 10957 scanned data points projected on two slices of the TESSM parameter space,  $m^2_{H_d}-m^2_{H_u}$ plane (Fig. \ref{stabps} left panel) 
and $\mu_D-B_D$ plane (Fig. \ref{mchHgg} left panel), with grey points being unstable against at least one among the UFB--1,2,3,3$'$ directions, while the colored ones satisfy all the stability constraints, 
Eqs.~(\ref{UFB1},\ref{UFB2},\ref{UFB3}). The color code,  shown in Fig.~\eqref{stabps} (right panel), represents the value of the triplet coupling $\lambda$ at the given data point. 
Besides the requirements for an unstable trivial vacuum state and positive scalar squared masses, which make large regions of parameter space inaccessible, the stability constraints evidently rule out 
the region featuring a large negative soft squared mass $m^2_{H_d}$ as well as a large $|\mu_D|$, with both parameters in absolute terms being smaller than 1 TeV$^2$ and TeV, respectively, for stable data points. This in turn limits the mass of the lightest chargino and neutralino. 

In Fig.~\eqref{mchHgg}, right panel, we plot the TESSM cross section relative to its SM value for a light Higgs boson, produced at LHC and then decaying into a diphoton, as a function of the lightest 
chargino mass. Also shown are the average value (solid line) measured by ATLAS and CMS \cite{Aad:2014eha,CMS:2014ega}, and the lower 1$\sigma$ bound (dashed line). 
The ratio of cross sections for the lightest Higgs boson, produced at LHC, decaying into $ij$ particles is defined by
\be
\hat{\mu}_{ij}=\frac{\sigma_{\textrm{tot}}{\textrm{Br}}_{ij}}{\sigma_{\textrm{tot}}^{\textrm{SM}} \textrm{Br} ^{\textrm{SM}}_{ij}}\ ,\quad \sigma_{\rm tot}=\sum_{\Omega=h,qqh,\ldots}\!\epsilon_\Omega\sigma_{\Omega} \ ,
\label{LHCb}\ee
where $\textrm{Br}_{ij}$ is the lightest Higgs boson branching ratio into the $ij$ particles, $\sigma_\Omega$ the production cross section of the given final state $\Omega$, and $\epsilon_\Omega$ 
is the corresponding efficiency. The details of the calculation of the branching ratio to diphoton and Higgs production total cross section in TESSM are given in \cite{Bandyopadhyay:2014tha}, and we do 
not repeat them here. From Fig.~\eqref{mchHgg} it is apparent that the stability constraints require the mass of the lightest chargino (and neutralino) to be lighter than about 700 GeV, for the scanned 
data points. As a consequence, within TESSM, a large deviation from the SM prediction is likely to be observed.

Since a lighter mass spectrum is usually associated in supersymmetric theories with  lower fine tuning (FT), it is interesting to look also at the effect of the stability constraints on the heavier stop mass together with the fine tuning at each scanned data point. A simple estimate of FT in supersymmetry is given by the logarithmic derivative of the EW vev $v_w$ with respect to a given model parameter $\mu_p$ \cite{Ellis:1986yg,Barbieri:1987fn}: this represents the change of $v_w$ for a 100\% change in the given parameter, as defined below:
\be
\mathbf{f}_{\mu_p}\equiv \frac{\partial  \log  v_w^2}{\partial  \log  \mu_p ^2 \left(\Lambda\right)  }\ ,\quad\mu_p ^2 \left(\Lambda\right) =\mu_p ^2 \left(M_Z\right)+\frac{\beta _{\mu_p ^2}  }{16 \pi ^2} \log   \left(\frac{\Lambda}{M_Z}\right)\ ,\quad
\beta _{\mu_p ^2}=16 \pi ^2 \frac{d\mu_p ^2}{d\text{logQ}}\ ,
\ee
where in parenthesis is the renormalisation scale of $\mu_p$. The FT in $m_{H_u}^2$, defined to be equal to $\mathbf{f}_{m_{H_u}} $, is then given by \cite{Bandyopadhyay:2014tha}:
\bea\label{FTdef}\nonumber
{\rm FT} &=&\frac{\log\left(\Lambda /M_Z\right)}{16\pi \partial_{v_w^2}m_{H_u}^2}\left(6 y_t^2 A_t^2+3 \lambda^2 A_T^2+3 \lambda ^2 m_{H_d}^2+3 \lambda ^2 m_T^2+3 \lambda ^2 m_{H_u}^2-2 g_Y^2  M_1^2-6 g_L^2 M_2^2\right.\\
 &+& 6 m_Q^2 y_t^2 + \left.6 m_{\tilde{t}}^2 y_t^2+6 m_{H_u}^2 y_t^2+ g_Y^2 \left(3 m_{\tilde{b}}^2-m_{H_d}^2-3 m_L^2+3 m_Q^2-6 m_{\tilde{t}}^2+m_{H_u}^2+3 m_{\tilde{\tau}}^2\right)\right) ,\nonumber\\
\eea
where the derivative in the denominator acts on the expression of $m_{H_u}^2$, Eqs.~\eqref{stabV}.
\begin{figure}[htb]
\centering
\includegraphics[width=0.46\textwidth]{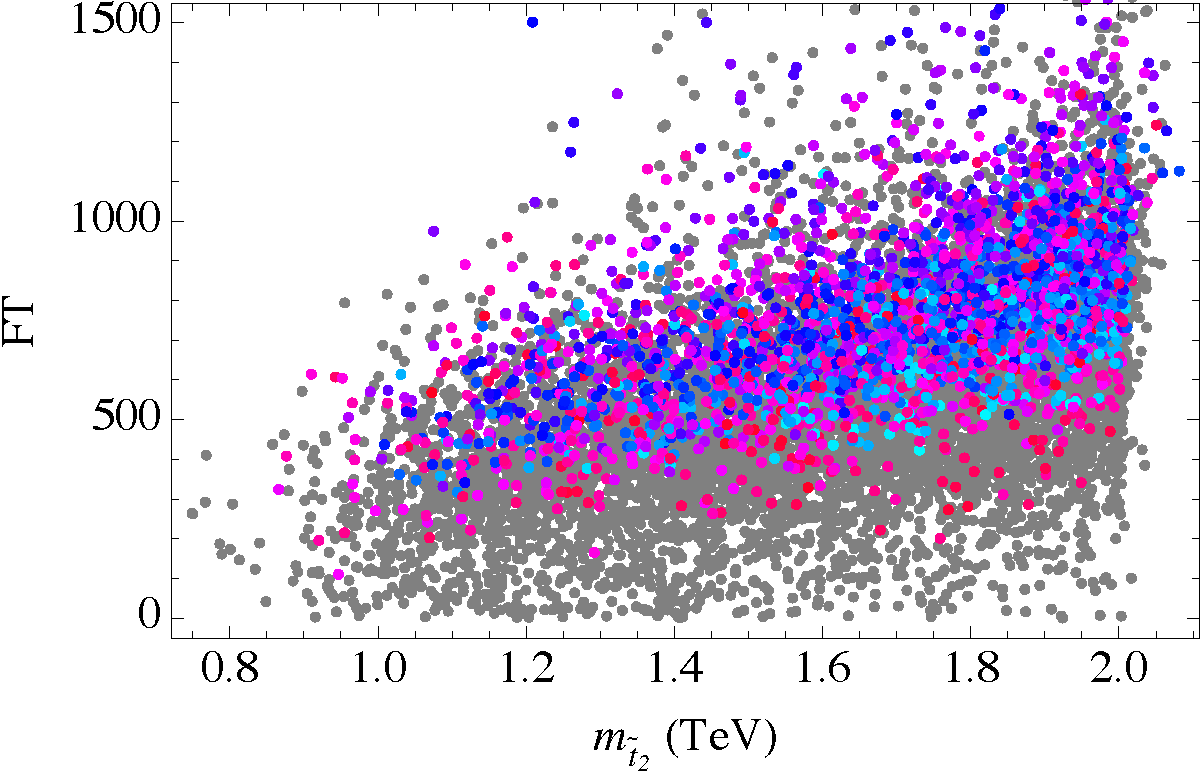}\hspace{0.1cm}
\includegraphics[width=0.46\textwidth]{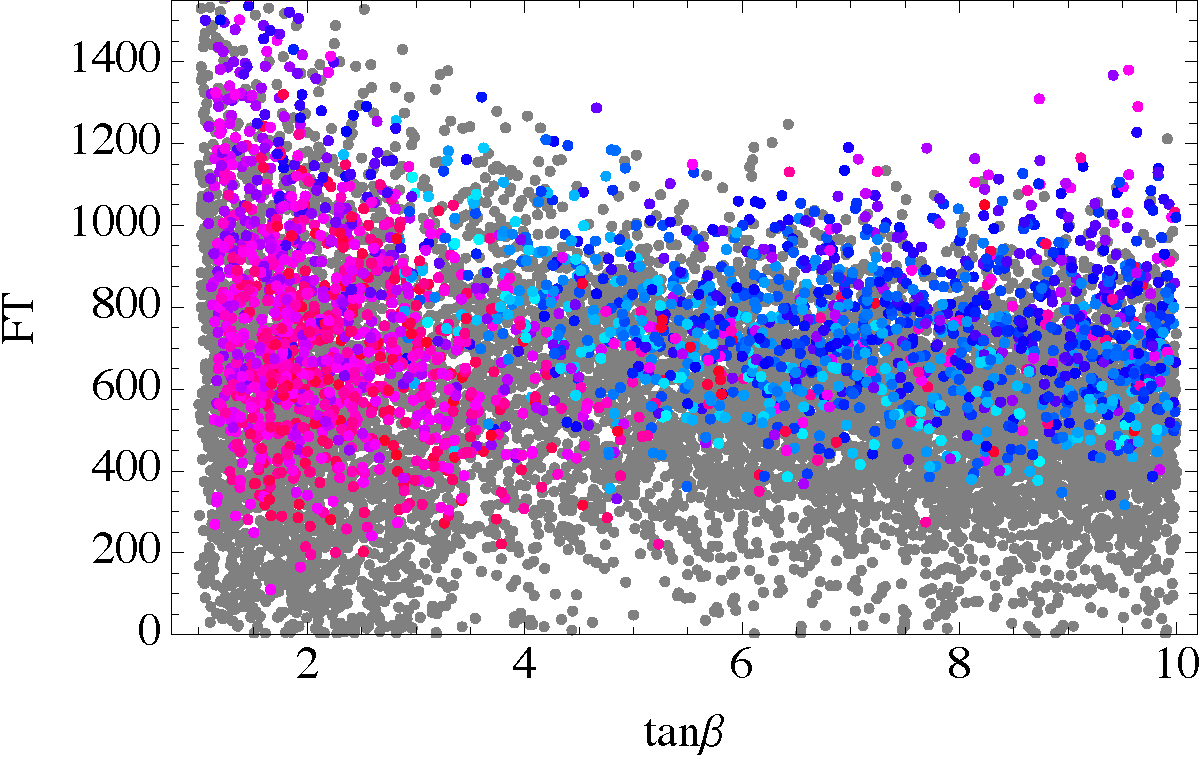}\hspace{0.1cm}
\caption{FT as a function of the heavier stop mass (left panel) and $\tan\beta$ (right panel). The grey points do not satisfy at least one of the stability constraints in 
Eqs.~(\ref{UFB1},\ref{UFB2},\ref{UFB3}), while the viable data points are colored according to the value of the triplet coupling $\lambda$ as shown in Fig.~\eqref{stabps} (right panel).}
\label{FTmsttb}
\end{figure}
In Fig.~\eqref{FTmsttb} we plot the scanned data points as a function of the heavier stop mass (left panel) and $\tan\beta$ (right panel). The grey points do not satisfy at least one of the stability 
constraints in Eqs.~(\ref{UFB1},\ref{UFB2},\ref{UFB3}), while the viable data points are colored according to the value of the triplet coupling $\lambda$ as shown in Fig.~\eqref{stabps} (right panel). 
An interesting (and unwelcome) effect of the stability constraints is that the points with the least amount of FT are actually ruled out, and as a result FT is on average 26\% higher after imposing the 
stability constraints, somewhat offsetting the advantage for naturality of the tree level triplet contribution to the MSSM Higgs boson mass. With increasing heavier stop mass the amount of FT on average increases as well, as expected, and a larger portion of data points satisfy the stability constraints. Indeed also for small values of $\tan\beta$, or equivalently large values of the triplet coupling $\lambda$, data points featuring a very low amount of FT are unstable against tunneling from the EW potential minimum to a vacuum along (at least one of) the UFB--1,2,3,3$'$ directions.


\section{Conclusions}\label{con}
In this paper we studied the Unbounded From Below (UFB) directions in the potential of a $Y=0$ SU$(2)$ triplet chiral superfield extension of MSSM, also called Triplet Extended Supersymmetric Standard Model (TESSM), and the associated stability constraints on the model's parameter space. After introducing the model, we systematically looked for sets of nonzero vevs that can cancel the quartic terms belonging to the D and F sectors of the TESSM tree level potential, under some rather general and reasonable simplifying assumptions, like chiral and charge symmetry of the vevs. We found four inequivalent sets of vevs that allow for UFB directions in the tree level potential, which we labeled UFB--1,2,3,4, respectively. One more UFB direction (UFB--3$'$) is obtained by simply switching the sbottom couplings and masses in the potential, defined along UFB--3, with the corresponding stau quantities. Among those directions, UFB--2,3,3$'$ turn out to be entirely equivalent to those already found in the MSSM potential \cite{Casas:1995pd}, in the sense that they do not contain any triplet contribution. For viable points in the TESSM parameter space, moreover, UFB--4 is lifted if UFB--1,2,3 are lifted as well, which makes the UFB--4 stability constraint irrelevant. The relevant stability constraints require the tree level potential evaluated at the EW minimum to be deeper than at any point along UFB--1,2,3,4,3$'$, with couplings and masses renormalized at a suitable scale minimizing the (neglected) one loop contributions. To carry out renormalization, we furthermore derived the one loop beta functions for the TESSM dimensional parameters, which were not given in past literature. 

To assess the relevance of the stability constraints for TESSM, we first scanned the TESSM parameter space and collected 10957 data points which produce the observed SM mass spectrum and satisfy direct 
search constraints on their superpartners and heavy Higgses, as well as EW precision parameter and perturbativity constraints. We then further tested the viability of these data points by checking how 
many of them satisfy the UFB--1,2,3,3$'$ stability constraints: among the (otherwise viable) 10957 data points, only 24\% turned out to be actually stable, with UFB--3$'$ giving the tighter constraint, 
which rules out 59\% of the scanned data points. Two of the parameters that get constrained the most by stability are the soft up (Higgs mass, $m_{H_d}$, and the Higgs doublets supersymmetric 
mass, $\mu_D$, 
both generally smaller than about 1~TeV for the data points featuring a stable EW minimum. As a consequence both the lightest chargino and neutralino turn out to be lighter than about 700~GeV for the same 
set of viable data points. While these are eventually too heavy for detection at LHC, one observable effect is that TESSM stable points feature on average a larger deviation from the SM predicted Higgs 
decay rate to diphoton than unstable points. We have shown furthermore that fine tuning within TESSM is on average 26\% higher after imposing the stability constraints, somewhat offsetting the advantage 
for naturality of the tree level triplet contribution to the MSSM Higgs boson mass. 

Finally, we conclude by highlighting the fact that the stability constraints described in this paper, by ruling out a large portion of the model's parameter space and affecting the superpartners mass spectrum, should be taken into account in any phenomenological study of the triplet extension of MSSM.

\section*{Acknowledgements}
This work was partially supported by the Academy of Finland, project 267842. MD acknowledges the hospitality of the Department of Theoretical Physics, Indian Association for the Cultivation of Science
during the course of this work. SR acknowledges the hospitality of the University of Helsinki and the Helsinki Institute of Physics during the final stages of this work. 

\newpage

\appendix


\section{Beta Functions}\label{apbeta}

The beta functions at two loops in TESSM for the Yukawa couplings $y_t$, $y_b$, $y_\tau$, $\lambda$, as well as for the gauge couplings $g_3,g_2=g_L,g_1=\sqrt{5/3}\,g_Y$ were already given in \cite{Bandyopadhyay:2014tha}, using the same superpotential and soft terms as in Eqs.~(\ref{SP},\ref{softV}), so there is no need to write them again here. The dimensionful couplings are $\mu_T$, $\mu_H$, $M_1$, $M_2$, $M_3$, $b_T$, $b_H$, $h_T$, $h_t$, $h_b$, $h_\tau$, as well as the squared soft mass parameters  $m^2_{\hat{T}}$, $m^2_{\hat{H}_u}$, $m^2_{\hat{H}_d}$, $m_{\hat{Q}}^2$, $m_{\hat{u}}^2$, $m_{\hat{d}}^2$, $m_{\hat{L}}^2$, $m_{\hat{e}}^2$. Their beta functions at one loop are defined by
\be
\frac{d z_x}{d t}=\frac{\beta _{z_x}^{(1)}}{16 \pi ^2}\quad {\rm for} \quad z_x=\mu_T,\mu_H,M_1,M_2,M_3,b_T,b_H,h_T,h_t,h_b,h_\tau \ ;\quad t=\log\frac{\Lambda}{\Lambda_{\rm EW}}\ ,
\ee
and by
\be
\frac{d z_x}{d t}=\frac{\beta _x^{(1)}}{16 \pi ^2}\quad {\rm for} \quad z_x=m^2_{\hat{T}},m^2_{\hat{H}_u},m^2_{\hat{H}_d},m_{\hat{Q}}^2,m_{\hat{u}}^2,m_{\hat{d}}^2,m_{\hat{L}}^2,m_{\hat{e}}^2 \ .
\ee
In the renormalization scheme using dimensional reduction (see  \cite{Martin:1993zk} and references therein) with modified minimal subtraction ($\overline{\rm DR}$) we find
\bea
\beta _{h_T}^{(1)}&=&-\frac{3}{5} g_1^2 h_T-7 g_2^2 h_T+3 h_T y_b^2+3 h_T y_t^2+h_T y_{\tau }^2+\frac{6}{5} g_1^2 M_1 \lambda _T+14 g_2^2 M_2 \lambda _T+6 h_b y_b \lambda _T+6
   h_t y_t \lambda _T\nonumber\\ &+&2 h_{\tau } y_{\tau } \lambda _T+24 h_T \lambda _T^2\ , \\
\beta _{h_t}^{(1)}&=&-\frac{13}{15} g_1^2 h_t-3 g_2^2 h_t-\frac{16}{3} g_3^2 h_t+h_t y_b^2+\frac{26}{15} g_1^2 M_1 y_t+6 g_2^2 M_2 y_t+\frac{32}{3} g_3^2 M_3 y_t+2 h_b y_b
   y_t+18 h_t y_t^2\nonumber\\ &+&6 h_T y_t \lambda _T+3 h_t \lambda _T^2\ , \\
\beta _{h_b}^{(1)}&=&-\frac{7}{15} g_1^2 h_b-3 g_2^2 h_b-\frac{16}{3} g_3^2 h_b+\frac{14}{15} g_1^2 M_1 y_b+6 g_2^2 M_2 y_b+\frac{32}{3} g_3^2 M_3 y_b+18 h_b y_b^2+2 h_t y_b
   y_t+h_b y_t^2\nonumber\\ &+&2 h_{\tau } y_b y_{\tau }+h_b y_{\tau }^2+6 h_T y_b \lambda _T+3 h_b \lambda _T^2\ , \\
\beta _{h_{\tau }}^{(1)}&=&-\frac{9}{5} g_1^2 h_{\tau }-3 g_2^2 h_{\tau }+3 h_{\tau } y_b^2+\frac{18}{5} g_1^2 M_1 y_{\tau }+6 g_2^2 M_2 y_{\tau }+6 h_b y_b y_{\tau }+12
   h_{\tau } y_{\tau }^2+6 h_T y_{\tau } \lambda _T+3 h_{\tau } \lambda _T^2  \ ,
\\
\beta _{\mu _T}^{(1)}&=&\mu _T \left(-8 g_2^2+4 \lambda _T^2\right)\ ,  \quad
\beta _{\mu _H}^{(1)}=\mu _H \left(-\frac{3}{5} g_1^2-3 g_2^2+3 y_b^2+3 y_t^2+y_{\tau }^2+6 \lambda _T^2\right)\ , \\
\beta _{M_1}^{(1)}&=&\frac{66}{5} g_1^2 M_1\ , \quad
\beta _{M_2}^{(1)}=6 g_2^2 M_2  \ , \quad
\beta _{M_3}^{(1)}=-6 g_3^2 M_3\ , \\
\beta _{b_T}^{(1)}&=&-8 g_2^2 b_T+4 b_T \lambda _T^2+16 g_2^2 M_2 \mu _T+8 h_T \lambda _T \mu _T\ , \\
\beta _{b_H}^{(1)}&=&-\frac{3}{5} g_1^2 b_H-3 g_2^2 b_H+6 h_b y_b \mu _H+3 b_H y_t^2+6 b_H \lambda _T^2+b_H y_{\tau }^2+3 b_H y_b^2+\frac{6}{5} g_1^2 M_1 \mu _H+6 g_2^2 M_2 \mu
   _H\nonumber\\ &+&6 h_t \mu _H y_t+12 h_T \mu _H \lambda _T+2 h_{\tau } \mu _H y_{\tau }\ .
\eea
By defining the quantity
\be
S=m_{\hat{H}_u}^2-m_{\hat{H}_d}^2+3 m_{\hat{Q}}^2-6 m_{\hat{u}}^2+3 m_{\hat{d}}^2-3 m_{\hat{L}}^2+3 m_{\hat{e}}^2\ ,
\ee
we can write the square mass parameters' beta functions as
\bea
\beta _{\hat{T}}^{(1)}&=&4 \lambda _T^2 m_{\hat{H}_d}^2-16 g_2^2 M_2^2+4 h_T \lambda _T \mu _T+4 h_T^2+4 \lambda _T^2 m_{\hat{H}_u}^2+4 m_{\hat{T}}^2 \lambda _T^2\ , \\
\beta _{\hat{H}_u}^{(1)}&=&\frac{3}{5} g_1^2 S+6
   \lambda _T^2 m_{\hat{H}_d}^2-\frac{6}{5} g_1^2 M_1^2-6 g_2^2 M_2^2+6 h_t^2+6 h_T^2+6 y_t^2 m_{\hat{H}_u}^2+6 \lambda _T^2 m_{\hat{H}_u}^2+6 m_{\hat{Q}}^2 y_t^2+6 m_{\hat{u}}^2
   y_t^2\nonumber\\ &+&2 m_{\hat{T}}^2 \lambda _T^2 \ ,\\
 \beta _{\hat{H}_d}^{(1)}&=&6 y_b^2 m_{\hat{H}_d}^2+6 y_b^2 m_{\hat{d}}^2+6 h_b^2+6 y_b^2 m_{\hat{Q}}^2+6 \lambda _T^2 m_{\hat{H}_d}^2+2 y_{\tau }^2 m_{\hat{H}_d}^2+2
   m_{\hat{e}}^2 y_{\tau }^2-\frac{6}{5} g_1^2 M_1^2-6 g_2^2 M_2^2-\frac{3}{5} g_1^2 S\nonumber\\ &+&2 h_{\tau }^2+6 h_T^2+6 \lambda _T^2 m_{\hat{H}_u}^2+2 m_{\hat{L}}^2 y_{\tau }^2+2 m_{\hat{T}}^2
   \lambda _T^2\ ,\\
 \beta _{\hat{Q}}^{(1)}&=&2 y_b^2 m_{\hat{H}_d}^2+2 y_b^2 m_{\hat{d}}^2+2 h_b^2+2 y_b^2 m_{\hat{Q}}^2-\frac{2}{15} g_1^2 M_1^2-6 g_2^2 M_2^2-\frac{32}{3} g_3^2
   M_3^2+\frac{1}{5}g_1^2 S+2 h_t^2+2 y_t^2 m_{\hat{H}_u}^2\nonumber\\ &+&2 m_{\hat{Q}}^2 y_t^2+2 m_{\hat{u}}^2 y_t^2\ , \\
\beta _{\hat{u}}^{(1)}&=&-\frac{32}{15} g_1^2 M_1^2-\frac{32}{3} g_3^2 M_3^2-\frac{4}{5} g_1^2 S+4 h_t^2+4 y_t^2 m_{\hat{H}_u}^2+4 m_{\hat{Q}}^2 y_t^2+4 m_{\hat{u}}^2 y_t^2\ ,\\
\beta _{\hat{d}}^{(1)}&=&4 y_b^2 m_{\hat{H}_d}^2+4 y_b^2 m_{\hat{d}}^2+4 h_b^2+4 y_b^2 m_{\hat{Q}}^2-\frac{8}{15} g_1^2 M_1^2-\frac{32}{3} g_3^2 M_3^2+\frac{2}{5} g_1^2 S\ ,\\
\beta _{\hat{L}}^{(1)}&=&2 y_{\tau }^2 m_{\hat{H}_d}^2+2 m_{\hat{e}}^2 y_{\tau }^2-\frac{6}{5} g_1^2 M_1^2-6 g_2^2 M_2^2-\frac{3}{5} g_1^2 S+2 h_{\tau }^2+2 m_{\hat{L}}^2
   y_{\tau }^2\ ,\\
\beta _{\hat{e}}^{(1)}&=&4 y_{\tau }^2 m_{\hat{H}_d}^2+4 m_{\hat{e}}^2 y_{\tau }^2-\frac{24}{5} g_1^2 M_1^2+\frac{6}{5} g_1^2 S+4 h_{\tau }^2+4 m_{\hat{L}}^2 y_{\tau }^2\ .
\eea



\begin{thebibliography}{99}
\bibitem{CMS:2014ega} 
  CMS Collaboration [CMS Collaboration],
and studies of the compatibility of its couplings with the standard model,''
  CMS-PAS-HIG-14-009.
\end{thebibliography}
\end{document}